\documentclass[a4paper,10pt]{article}
\usepackage[affil-it]{authblk}
\usepackage{amsmath}
\usepackage{amssymb}
\usepackage{hyperref}
\usepackage{geometry}
\usepackage{color}
\usepackage{graphicx}
\usepackage{subfig}
\usepackage{mwe}
\usepackage[font=small,labelfont=bf]{caption}
\usepackage{lineno}
\usepackage[T1]{fontenc}
\usepackage[latin9]{inputenc}
\usepackage{amstext}
\usepackage{babel}
\usepackage{booktabs}
\usepackage{verbatim}

\begin{document}

\title{The divergence and curl in arbitrary basis}
\author{ Waleska P. F. de Medeiros, Rodrigo R. de Lima, 
Vanessa C. de  Andrade, Daniel M\"uller\thanks{muller@fis.unb.br} \\
{\small Instituto de F\'\i sica, Universidade de Bras\'\i lia,}\\{\small Bras\'\i lia, DF}}

\maketitle

\begin{abstract}
In this work, the divergence and curl operators are obtained using the coordinate free non rigid basis formulation of differential geometry. Although the authors have attempted to keep the presentation self-contained as much as possible, some previous exposure to the language of differential geometry may be helpful. In this sense the work is aimed to late undergraduate or beginners graduate students  interested in mathematical physics. To illustrate the development, we graphically present the eleven coordinate systems in which the Laplace operator is separable. We detail the development of the basis and the connection for the cylindrical and paraboloidal coordinate systems. We also present in \cite{RAM} codes both in Maxima and Maple for the spherical orthonormal basis, which serves as a working model for calculations in other situations of interest. Also in \cite{RAM} the codes to obtain the coordinate surfaces are given.
\end{abstract}

\newline

\global\long\def\imsize{0.83\columnwidth}
 \global\long\def\halfsize{0.45\columnwidth}
\global\long\def\big{1.2\columnwidth}
\global\long\def\peq{0.3\columnwidth}

\section{Introduction}
At the undergraduate level, it can be noticed that in the majority of the books on classical mechanics \cite{Symon}, and also in electromagnetism \cite{Griffiths}, use is made of the famous unit vectors $\hat{r}$, $\hat{\theta},$
$\hat{\phi}$, or  $\hat{\rho}$, $\hat{\theta},$ $\hat{z}$, respectively in the spherical and cylindrical cases. Both these bases are called non coordinate. In contrast, the cartesian basis $\hat{x}$, $\hat{y}$ and $\hat{z}$ is called a coordinate basis.  We will get to this point in section \ref{2}.

The intention of this work is to provide a coordinate independent expression for the vector operators, imposing the orthonormality condition. The subject is not new as we will discuss just in the following. It is a formalism, such that it will result in the same equations, written in a different form. We also present the 11 coordinate systems for which Laplace 
operator is separable \cite{Morse_Feshbach}. 

On the other hand, vector calculus, as we know it, was developed by Josiah W. Gibbs in Yale in the second half of the XIX century which ended up with the famous book, {\it Vector Analysis} written together with Edwin B. Wilson in 1901, \cite{VA}. Gibbs's book only addresses coordinate basis. 

Sommerfeld's book \cite{Sommerfeld} is the older reference that we know of, in which the vector operators in curvilinear systems are presented. In this book, the operators are obtained through their integral definition. Circulation of a vector $\vec{V}$ in a closed loop $\gamma$, $\oint_\gamma \vec{V}.d\vec{r}$, is used to obtain the curl $\nabla \times \vec{V}$. And flux over a closed surface $\Sigma$, $\oint_\Sigma \vec{V}.d\vec{\sigma}$ to obtain the divergence $\nabla .\vec{V}$. All other textbooks on mathematical physics in which we were able find the vector operators in curvilinear systems  \cite{Hassani,Hobson,Arfken68,Arfken13,Boas,Morse2}, follow the same technique as in Sommerfeld's book.

In this present work, there is no intention to develop a systematic review in differential geometry. Anyway, in order to have a mostly self contained text, we shall briefly introduce the concepts of tensors and the covariant derivative. Many modern textbooks on mathematical physics do present differential geometry as a standard subject in which the interested reader can look into, for example \cite{Hassani,Hobson,Arfken13}. 

Looking into many General Relativity (GR) textbooks, it is possible to find the use non coordinate orthonormal basis, which in the GR context are known as tetrads, see for example \cite{Stephani,Felice}. In \cite{Dray}, the vector operators are obtained using differential forms. For example, in \cite{MTW} page 213 in exercise 8.6, it is presented the divergence of a vector field in spherical coordinates using the same technique which we are presenting here in our work. In some sense, the examples carried out by us complement the ones in  the book of Misner, et al. \cite{MTW}, since both the divergence and curl of a vector field are developed here.

The subject is not new, as it is well known that attached to each curve there's a locally orthogonal frame which was independently discovered by Jean F. Frenet  and Joseph A. Serret, respectively in 1847 and 1851. During the XIX century, Jean G. Darboux generalized the Frenet-Serret frame to surfaces, developing the {\it tr\`\i edre mobile} which culminated with the four volumes  published during the years 1887-1896, \cite{Darboux}. Moving frames were latter addressed by \'Elie Cartan in connection to Lie Groups in 1937, \cite{Cartan}. The equations governing Darboux frame, in modern language, are called Cartan structure equations, see for example \cite{wiki}. 

We must mention that there is a very interesting previous work on divergence and curl for electromagnetism for some particular cases \cite{Dray2}.  As an application of the formalism, we show analytically how to obtain the divergence and curl of a vector for the cylindrical and a non trivial example, the paraboloidal orthonormal basis. We also present in \cite{RAM}, codes both in Maxima \cite{maxima}, \cite{manuais} and Maple \cite{maple} to obtain the operators for the spherical orthonormal basis, and also the codes to obtain the coordinate surfaces. It is very easy to adapt the code to the other coordinate systems, or to any other situation of interest. 
The work is aimed at the second half of the physics course or beginning graduate students. The subject could be easily mastered to anyone interested to include this formalism in a ordinary course of mathematical physics to be given in the classroom to the students. 

The article is organized as follows: in section \ref{2} the coordinate free formulation is developed; also in this section, the appropriate connection for an arbitrary basis is given. In section \ref{3} we list a set of coordinates for which Laplace operator is separable, and we show how to find the appropriate orthonormal basis adapted for each frame. The divergence and curl of a vector for the cylindrical, and paraboloidal orthonormal bases are also presented in this section. Our final remarks are presented in the conclusions. 

\section{Coordinate free formulation\label{2}}
\subsection{Bases}

We begin the discussion by first clarifying some facts about bases through an example, after which getting more formal. Despite the attempt to keep the text self-consistent, prior knowledge in differential geometry may be useful.
\newline
\begin{figure}[ht]
\begin{center}
\begin{tabular}{c}
{\includegraphics[width=0.35\textwidth]{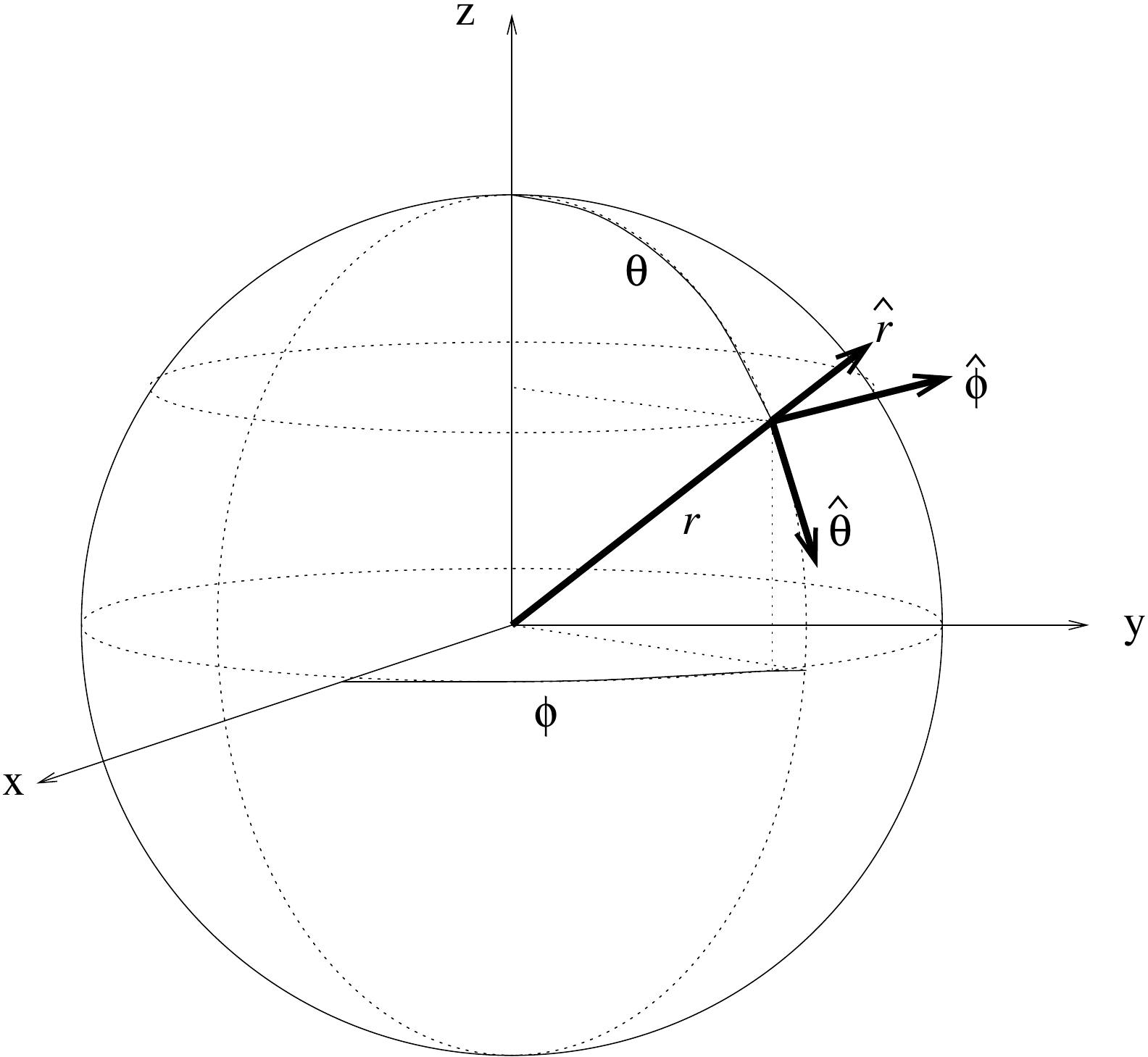}}
\end{tabular}
\end{center}
\caption{Spherical coordinates, it is shown the orthonormal physical basis, $\hat{r}$, $\hat{\theta}$ and $\hat{\phi}$. 
 }  
\end{figure}
The Figure 1 shows the very example of the orthonormal unit basis vectors important to physics. It can be seen there, that all the basis vectors get modified \footnote{Except for $\hat{\theta}$ at the equator and $\hat{r}$ at the poles, which do not depend on the coordinate $\phi$} under translations in $\phi$. While $\hat{r}$ and $\hat{\theta}$ depend on $\theta$ and all the basis vectors do not depend in modifications in $r$. Therefore, only the vector $\hat{r}$ can be integrated in the same sense of a velocity vector to give the coordinate line $r$. It is not possible to integrate the other two  basis vector. This is an example of an orthonormal non coordinate basis. It is not possible to integrate all the coordinate basis vectors to give rise to the coordinate lines.

The intention of this work is to develop the vector operators for arbitrary orthonormal non coordinate bases. We choose a very traditional path, introducing first coordinate bases and then arbitrary orthonormal bases. There is a strong reason for this, since restricting to Euclidean space, it is always possible to choose the cartesian basis, which is the sole example in which a coordinate basis is also orthonormal.Both bases can be used for spherical coordinates, being the orthonormal the more important to physics.

Now we turn to the general case, such that latin  indices $i,\,j,\,k,\, etc.$ run from 1 to 3 and concern coordinate basis, while latin indices $a,\,b,\,c,\,etc$ label each orthonormal vector.

For us, an arbitrary vector $V$ is always connected to the directional derivative, in the sense that

\begin{equation}
V^{i}\partial_{i}f(x)=f_{|V}\label{vector}
\end{equation}

where $\partial_i=\frac{\partial}{\partial x^i}$ are the partial derivative operators in the coordinate directions, which, when interpreted as vectors, form the coordinate basis. Here $x\in E^3$ is point in Euclidean space, $f(x)$ is any scalar function and it is used the usual convention of summation over repeated indices. Therefore, a vector $V$ can be written as
\begin{equation}
V=V^{i}\partial_{i}.
\end{equation}

Now we define a change of basis. Starting with any of the coordinate  bases, we change to an arbitrary basis $e_{a}$

\begin{eqnarray}
 &  & \partial_{i}=M_{i}{}^{a}e_{a}\nonumber\\
 &  & e_{a}=\left(M^{-1}\right)_{a}^{\,j}\partial_{j},
\end{eqnarray}
where necessarily the basis transformation matrix $M_{i}{}^{a}$
must be invertible and cover all the space, except for sets of measure zero.

Therefore in an arbitrary basis, the components of a vector transform as  
\begin{equation}
V=V^{i}\partial_{i}=V^{i}M_{i}{}^{a}e_{a}=\tilde{V}^{a}e_{a},
\end{equation}
where
\begin{equation}
\tilde{V}^{a}=V^{j}M^{a}{}_{j}.
\end{equation}

Coordinate transformations are particular cases of basis transformations. Suppose we have a coordinate transformation
\begin{eqnarray}
 &  & x^{i}=x^{i}(\tilde{x})\nonumber\\
 &  & \tilde{x}^{i}=\tilde{x}^{i}(x).
\end{eqnarray}
Due to the chain rule  
\begin{eqnarray}
 &  & \frac{\partial}{\partial x^{i}}=\frac{\partial\tilde{x}^{j}}{\partial x^{i}}\frac{\partial}{\partial\tilde{x}^{j}}\nonumber\\
 &  & M_{i}{}^{j}=\frac{\partial\tilde{x}^{j}}{\partial x^{i}},
\end{eqnarray}
so that the transformation matrix is the Jacobi matrix. 

Now we define the commutator between to vectors, say $A$ and $B$ 
\begin{eqnarray}
 &  & [A,B]=A^{i}\partial_{i}\left(B^{j}\partial_{j}\right)-B^{i}\partial_{i}\left(A^{j}\partial_{j}\right)\nonumber\\
 &  & =A^{i}\left(\partial_{i}B^{j}\right)\partial_{j}-B^{i}\left(\partial_{i}A^{j}\right)\partial_{j},
\end{eqnarray}
this last line follows from the fact that when $[A,B]$ is applied to differentiable functions, $\partial_{i}\partial_{j}f(x)=\partial_{j}\partial_{i}f(x).$ Therefore the commutator between two vectors must be itself also a vector, 
\begin{equation}
[A,B]^{j}=A^{i}\left(\partial_{i}B^{j}\right)-B^{i}\left(\partial_{i}A^{j}\right).
\end{equation}

The commutator between two vectors measure their independency and a basis is called coordinate if all the commutators between them vanish, for instance
\begin{equation}
[\partial_{i},\partial_{j}]=0.
\end{equation}
For example, in the cartesian orthonormal case, we have $\hat{x}=1\partial_{x}+0\partial_{y}+0\partial_{z}=\partial_{x}$,
$\hat{y}=0\partial_{x}+1\partial_{y}+0\partial_{z}=\partial_{y}$
and $\hat{z}=0\partial_{x}+0\partial_{y}+1\partial_{z},=\partial_{z}$
and we can easily convince ourselves that all the commutators between $\hat{x}\,,\hat{y}$ and $\hat{z}$ are zero.

\subsection{Metric and connection}
In this section the concepts of metric and the covariant derivative are briefly introduced, so that we have the appropriate geometrical operators necessary to obtain the divergence and curl. For a much more complete and detailed approach on differential geometry, see for example \cite{Hassani}.

The orthonormal basis can be expanded with respect to the coordinate basis $\partial_i$, $e_a=e_{\;a}^i\partial_i$, such that the $i$ enumerates the components of the vector and $a$ labels each of the basis element. 
There is no intention to emphasize differential forms in this text, so we just present the dual basis $\omega^{b}_{\,\,i}$ such that $\omega^{b}_{\,\,i}e_{\,a}^i=\delta_{a}^{b}$
and the corresponding coordinate basis $dx^{j}$, such that the scalar product is $dx^{i}\partial_{j}=\delta_{j}^{i}.$ 

We will define the connection through the metric tensor, although we know this is not a necessary condition. Only Euclidean geometry is going to be addressed in this present work, which in cartesian coordinates reproduces Pythagora's theorem 
\begin{equation}
ds^{2}=dx^{2}+dy^{2}+dz^{2}=g_{ij}dx^{i}dx^{j}.
\end{equation}
Therefore, the Euclidean metric is \footnote{Remind that only in Euclid's geometry, the cartesian coordinate basis coincides with the orthonormal basis, such that $g_{ij}=g_{ab}=\delta_{ab}$ which is the identity.}
\begin{equation}
g_{ij}=\left(\begin{array}{ccc}
1 & 0 & 0\\
0 & 1 & 0\\
0 & 0 & 1
\end{array}\right).
\end{equation}
The line element $ds^2$ turns into  
\begin{eqnarray}
 &  & ds^{2}=\left(\frac{\partial x}{\partial\tilde{x}^{1}}d\tilde{x}^{1}+\frac{\partial x}{\partial\tilde{x}^{2}}d\tilde{x}^{2}+\frac{\partial x}{\partial\tilde{x}^{3}}d\tilde{x}^{3}\right)^{2}+\left(\frac{\partial y}{\partial\tilde{x}^{1}}d\tilde{x}^{1}+\frac{\partial y}{\partial\tilde{x}^{2}}d\tilde{x}^{2}+\frac{\partial y}{\partial\tilde{x}^{3}}d\tilde{x}^{3}\right)^{2}\nonumber\\
 &  & +\left(\frac{\partial z}{\partial\tilde{x}^{1}}d\tilde{x}^{1}+\frac{\partial z}{\partial\tilde{x}^{2}}d\tilde{x}^{2}+\frac{\partial z}{\partial\tilde{x}^{3}}d\tilde{x}^{3}\right)^{2}=\tilde{g}_{ij}d\tilde{x}^{i}d\tilde{x}^{j}, 
\end{eqnarray}
in an arbitrary coordinate basis $d\tilde{x}^i,$ such that the components of the metric transform as 
\begin{equation}
\tilde{g}_{ij}=\frac{\partial x^{m}}{\partial\tilde{x}^{i}}\frac{\partial x^{n}}{\partial\tilde{x}^{j}}g_{mn}. \label{trmt}
\end{equation}
  As we said before the matrix $\frac{\partial x^{m}}{\partial\tilde{x}^{i}}$is called Jacobian since it involves the change between two coordinate bases.
First by a coordinate change the components of the metric transform to $\tilde{g}_{ij}$ as (\ref{trmt}). After the coordinate change, the orthonormal basis is the one that brings the $\tilde{g}_{ij}$ given in (\ref{trmt}) to the Euclidean metric
\begin{equation}
g_{ab}=\left(\begin{array}{ccc}
1 & 0 & 0\\
0 & 1 & 0\\
0 & 0 & 1
\end{array}\right)=\tilde{g}_{ij}e^{i}{}_{a}e^{j}{}_{b}.
\label{base}
\end{equation}
This is the orthonormal basis chosen, such that the tensor and vector components in both bases are related as $V^i=V^ae^i{}_a$.

The ordinary partial derivative  
\begin{equation}
\partial_{i}(V^{j}\partial_j), \label{dparcial}
\end{equation}
is not invariant through a change of basis, because there will be necessarily the derivative of the basis transformation matrix 
 $\left(M^{-1}\right)^{j}_{\,a}$ associated to the components $V^{j}=\left(M^{-1}\right)_{\,a}^{j}\tilde{V}^{a}$. 
 
The derivative $\partial_i$ in (\ref{dparcial}) must be modified to achieve the desired invariance, which is called the covariant derivative $\nabla_i$. Necessarily the covariant derivative of the basis vector $\partial_i$ must be a linear combination of the basis vectors, 
\begin{equation}
\nabla_i\partial_j=\Gamma^k_{\,ji}\partial_k.\label{Christoffel}
\end{equation}
Keeping in mind the above equation (\ref{Christoffel}), (\ref{dparcial}) must be modified to

\begin{equation}
\nabla_i( V^j\partial_j)=(\partial_iV^j+ \Gamma^j_{\,ki} V^k) \partial_j, \label{covariant_d}
\end{equation} 
which satisfies the linearity condition
\begin{equation}
\nabla_i(V^j+A^j)=\nabla_i(V^j)+\nabla_i(A^j).
\end{equation}
Now it is possible to define the the covariant derivative $\nabla_{a}$, as an operator acting on vectors. This covariant derivative is the appropriate one that must be used in order to obtain the divergence and curl, which is the purpose of this work. First the covariant derivative $\nabla_a$ 
is the projection of the derivative $\nabla_i$ into the arbitrary basis $e_a^i$
\begin{equation}
\nabla_a\equiv e^i{}_a\nabla_i,
\end{equation}
where in order to shorten notation we write $\nabla_a$ instead of $\nabla_{e_a}$. Then, the covariant derivative of a scalar function coincides with the directional derivative
\begin{equation}
\nabla_{a}f=f_{|a}=e^{i}{}_{a}\partial_{i}f,\label{DC_E}
\end{equation}
where again in order to shorten notation and in accordance with the directional derivative (\ref{vector}), we write $|a$ instead of $|e_a$. While the covariant derivative of the product of a scalar and a vector $fV^i$ satisfies the product rule 
\begin{equation}
\nabla_i(fV^j)=\partial_i(f) V^j+f\nabla_i(V^j),\label{D(fV)}
\end{equation}
where $\nabla_i(V^j)$ is given by (\ref{covariant_d}).
Recall that $V^j=e^j{}_a V^a$, such that $V^j$ is a linear combination of vectors labeled by $a$, $e^i{}_a$ multiplied by scalars $V^a$ so that according to (\ref{D(fV)})
\begin{equation}
\nabla_i(V^ae^j{}_{a})=e^j{}_a\partial_iV^a+V^a\nabla_ie^j{}_a. \label{D_iV^j}
\end{equation}
Now remind that $e^i{}_b\nabla_i=\nabla_b$ so that the projection of (\ref{D_iV^j}) gives
\begin{equation}
\nabla_b (e^j{}_a V^a)= \left[e^j{}_a e^i{}_b\partial_i (V^a)+V^a\nabla_b(e^j{}_a)\right].\label{operador}
\end{equation}

Again, necessarily the covariant derivative of an element of the basis, must be a linear combination of the elements of the basis 
\begin{equation}
\nabla_{b}e^j{}_{a}= \Gamma^{c}{}_{ab} e^j{}_{c},
\end{equation}
which results in the definition of the operator $\nabla_b$ acting on a vector $V^a$ as the $a-$component of (\ref{operador})
\begin{equation} 
\nabla_bV^a =(V_{|b}^{a}+ \Gamma^{a}{}_{cb} V^{c})
\label{derivada}.
\end{equation}

By applying the covariant derivative to the scalar product $A^{a}B_{a}$ results in the covariant derivative acting on a covariant vector
 \begin{equation}
\nabla_{a}B_{b}=B_{b|a}-\Gamma^{c}{}_{ba}B_{c}.
\end{equation}

The covariant derivative generalizes the parallel transport of a vector $A^a$ along a direction $B^a$
\begin{equation}
\frac{D}{D\theta}A^a=B^c\nabla_c A^a, 
\end{equation}
where $\theta$ is a parameter along the vector $B^a$, \cite{Hassani}. 

 We have the following two properties:
\begin{itemize}
\item metricity $\nabla_{c}g_{ab}=0$ which states that the metric is invariant under parallel transport 
\begin{eqnarray}
 &  & \nabla_{c}g_{ab}=g_{ab|c}-\Gamma^{d}{}_{ac}g_{db}-\Gamma^{d}{}_{bc}g_{ad} \label{cond1}\\
 &  & \Gamma_{a\,bc}+\Gamma_{b\,ac}=g_{ab|c}=0\label{cond2}, 
\end{eqnarray}
since our basis is orthonormal (\ref{base}), $g_{ab}=\text{diag}[1,1,1]$, all its directional derivatives vanish $g_{ab|c}=0$.

\item and zero torsion, which means that the commutator of two vectors coincide with their deficit upon interchange in parallel transport 
\begin{equation}
 \left(\Gamma^{c}{}_{ba}-\Gamma^{c}{}_{ab}\right)e_{c}=[e_{a},e_{b}]=D^{c}{}_{ab}e_c.
\end{equation}
If the basis $e^i{}_a$ is coordinate, all the commutators must vanish, and we recover that the connection in the coordinate basis $\Gamma^i{}_{jk}=\Gamma^i{}_{kj}$ is symmetric in the lower indices.
\end{itemize}
The above two items completely define the Levi-Civita connection \footnote{Connection (\ref{Gamma}) is strictly valid only when the basis is orthonormal (\ref{base}). Anyway, it's not difficult to generalize (\ref{Gamma}) when (\ref{cond2}) is different from zero.}
\begin{equation}
\Gamma_{a\,bc}=\frac{1}{2}\left(D_{a\,cb}-D_{b\,ca}-D_{c\,ba}\right). \label{Gamma}
\end{equation}

The indices are raised and lowered using Euclide's metric $g_{ab}=g^{ab}=\delta^{ab}$ which is the identity, for example, raising the $d$ index in (\ref{Gamma}) results in 
\begin{equation}
\Gamma^a{}_{bc}=\frac{1}{2}\left(D_{d\,cb}-D_{b\,cd}-D_{c\,bd}\right)\delta^{ad}=\frac{1}{2}\left(D^a{}_{cb}-D_{b\,c}^{\,\,\,\,\,a}-D_{c\,b}^{\,\,\,\,\,a}\right).
\label{conexao}
\end{equation}

Therefore, the commutators between the elements of the basis define the coefficients $D^a{}_{bc}$ which give raise to the connection. The covariant derivative, $\nabla_a$, finally define the divergence and curl of a vector 
\begin{equation}
\nabla_{a}V^{a}=V_{|a}^{a}+\Gamma^{a}{}_{ba}V^{b}
\end{equation}
\begin{eqnarray}
&&[\nabla\times V]^c=\frac{1}{2}\epsilon^{cab}\left( \nabla_aV_b -\nabla_bV_a\right)\nonumber\\
&&\nabla_{a}V_{b}-\nabla_{b}V_{a}=V_{b|a}-V_{a|b}-D^d{}_{ab}V_{d}, 
\end{eqnarray}
where $\epsilon^{cab}$ is Levi-Civita's skew symmetric tensor. 

\section{Different orthonormal basis \label{3}} 
The intention now is to apply the preceding formalism to obtain the divergence and curl in orthonormal basis for other coordinate systems. 
There are $11$ coordinate systems in which is possible to separate variables for the Laplace operator in Euclidean space \cite{Morse_Feshbach}. 

Laplace operator appears in hyperbolic elliptic and parabolic differential equations, which comprise most of the cases of interest in mathematical physics. 

As an example we work out analytically in details the cylindrical coordinates, and also the paraboloidal system as a non trivial case. The same proceeding for the spherical coordinates is presented as codes both in Maple and Maxima in \cite{RAM}. The other coordinates systems are presented and also shown graphically, with the intention that the reader himself can adapt the codes in Maple or Maxima and carry out the divergence and curl for each one of them. 

\subsection{Cylindrical coordinates \label{coord.cil}} 
As an example we choose the cylindrical coordinates 
\begin{eqnarray}
 &  & x=\rho\cos\theta\nonumber\\
 &  & y=\rho\sin\theta\nonumber\\
 &  & z=z.
\end{eqnarray}
The line element is the following
\begin{equation}
ds^{2}=(dx)^{2}+(dy)^{2}+(dz)^{2}=d\rho^{2}+\rho^{2}d\theta^{2}+dz^{2}.
\end{equation}
In this section the indices $\rho$, $\theta$ and $z$ are to be understood as the labeling of each one of the basis elements. With this in mind, the orthonormal basis is
\begin{eqnarray}
 &  & e_{\rho}=(1,0,0) =1\partial_\rho+0\partial_\theta+0\partial_z=\partial_\rho \nonumber\\
 &  & e_{\theta}=(0,1/\rho,0)=0\partial_\rho+\frac{1}{\rho}\partial_\theta+0\partial_z=\frac{1}{\rho}\partial_\theta\nonumber\\
 &  & e_{z}=(0,0,1)=0\partial_\rho+0\partial_\theta+1\partial_z=\partial_z.
\end{eqnarray}
The only non null commutator is the following 
\begin{eqnarray}
 &  & [e_{\rho},e_{\theta}]=(1\partial_{\rho}+0\partial_{\theta}+0\partial_{z})(0,1/\rho,0)=(0,-1/\rho^{2},0)\\
 &  & [e_{\rho},e_{\theta}]=-\frac{1}{\rho} e_{\theta}=D^{\theta}{}_{\rho\theta}e_{\theta}\\
 &  & D^{\theta}{}_{\rho\theta}=-\frac{1}{\rho}
 \end{eqnarray}
Resulting in the only non null component
\begin{eqnarray}
 &  & \Gamma_{\rho\,\theta\theta}=\frac{1}{2}\left(D_{\rho\,\theta\theta}-D_{\theta\,\theta\rho}-D_{\theta\,\theta\rho}\right)\\
 &  & \Gamma_{\rho\,\theta\theta}=-D_{\theta\,\theta\rho}=-\frac{1}{\rho},\\
 &  & \Gamma_{\theta\:\rho\theta}=D_{\theta\,\theta\rho}=\frac{1}{\rho}.
\end{eqnarray}
The divergence and curl are
\begin{eqnarray}
 &  & \nabla_{a}V^{a}=V_{|a}^{a}+\Gamma^{a}{}_{ba}V^{b}=V_{|a}^{a}+\frac{1}{\rho}V^{\rho}\nonumber\\
 &  & \nabla_{a}V_{b}-\nabla_{b}V_{a}=V_{b|a}-V_{a|b}-\left(\Gamma^{c}{}_{ba}-\Gamma^{c}{}_{ab}\right)V_{c}\nonumber\\
 &  & \left[\nabla_{a}V_{b}-\nabla_{b}V_{a}\right]_{\rho}=V_{z|\theta}-V_{\theta|z}\\
 &  & \left[\nabla_{a}V_{b}-\nabla_{b}V_{a}\right]_{\theta}=V_{\rho|z}-V_{z|\rho}\nonumber\\
 &  & \left[\nabla_{a}V_{b}-\nabla_{b}V_{a}\right]_{z}=V_{\theta|\rho}-V_{\rho|\theta}-(\Gamma^{\theta}{}_{\theta\rho}-\Gamma^{\theta}{}_{\rho\theta})V_{\theta}\nonumber\\
 &&=V_{\theta|\rho}-V_{\rho|\theta}+\frac{1}{\rho}V_{\theta}.
\end{eqnarray}

Here, for instance,  $V_{z|\theta}=e^i{}_\theta\partial_iV_z=e^{\theta}{}_{\theta}\partial_{\theta}V_z=\frac{1}{\rho}\partial_{\theta}V_z.$

\subsection{Paraboloidal Coordinates}
These coordinates can be defined either through Jacobi elliptic functions or through more simple functions, see for example \cite{Moon}
\begin{eqnarray}
&&x=\pm\sqrt{\frac{(a-\lambda)(a-\mu)(a-\nu) }{b-a}}\\ 
&&y=\pm\sqrt{\frac{(b-\lambda)(b-\mu)(b-\nu) }{a-b}}\\
&&z=\frac{1}{2}(a+b-\lambda-\nu -\mu), \label{coord_paraboloidais}
\end{eqnarray}
where $\lambda<b<\mu<a<\nu$. The entire space is covered using the $\pm$ signs. 
The advantage of using Jacobi elliptic functions is that their arguments corresponds to generalized angles. We used the maple code in \cite{RAM} with the coordinates defined in (\ref{coord_paraboloidais}). The metric in the coordinate basis reads 
\begin{equation}
g_{ab}=\left(\begin{array}{ccc}
\frac{(\lambda-\nu)(\lambda-\mu)}{4(\lambda-a)(\lambda-b)} & 0 & 0\\
0 & \frac{-(\lambda-\mu)(\mu-\nu)}{4(\mu-a)(\mu-b)} & 0\\
0 & 0 & \frac{(\nu-\lambda)(\nu-\mu)}{4(\nu-a)(\nu-b)}
\end{array}\right)=
\left(\begin{array}{ccc}
g_1 & 0 & 0\\
0 & g_2 & 0\\
0 & 0 & g_3
\end{array}\right)
\label{gi}
\end{equation}
The orthonormal basis vectors are 
\begin{eqnarray}
&&e_\lambda=1/\sqrt{g_1}\partial_\lambda\\
&&e_\mu=1/\sqrt{g_2}\partial_\mu\\
&&e_\nu=1/\sqrt{g_3}\partial_\nu
\end{eqnarray}
Their commutators are
\begin{eqnarray}
&&[e_\lambda,e_\mu]=-\frac{e_\lambda}{2(\lambda-\mu)\sqrt{g_2}}-\frac{e_\mu}{2(\lambda-\mu)\sqrt{g_1}}\\
&&[e_\lambda,e_\nu]=-\frac{e_\lambda}{2(\lambda-\nu)\sqrt{g_3}}-\frac{e_\nu}{2(\lambda-\nu)\sqrt{g_1}}\\
&&[e_\mu,e_\nu]=-\frac{e_\mu}{2(\mu-\nu)\sqrt{g_3}}-\frac{e_\nu}{2(\mu-\nu)\sqrt{g_2}}.
\end{eqnarray}
The divergence is 
\begin{eqnarray}
&&\nabla_a A^a=\frac{\partial_\lambda A^\lambda}{\sqrt{g_1}}+\frac{\partial_\mu A^\mu}{\sqrt{g_2}}+\frac{\partial_\nu A^\nu}{\sqrt{g_3}}+\frac{(2\lambda-\mu-\nu)}{2(\lambda-\nu)(\lambda-\mu)\sqrt{g_1}}A^\lambda\nonumber\\
&&+\frac{(\lambda+\nu-2\mu)}{2(\mu-\nu)(\lambda-\mu)\sqrt{g_2}}A^\mu-\frac{(\lambda+\mu-2\nu)}{2(\lambda-\nu)(\mu-\nu)\sqrt{g_3}} A^\nu,
\end{eqnarray}
where the $g_i$ are defined in (\ref{gi}). 
The curl in components is given by 
\begin{eqnarray}
&&[\nabla\times\vec{A}]^\lambda=\frac{\partial_\mu A^\nu}{\sqrt{g_2}}-\frac{\partial_\nu A^\mu}{\sqrt{g_2}}+\frac{A^\mu}{2(\mu-\nu)\sqrt{g_3}}+\frac{A^\nu}{2(\mu-\nu)\sqrt{g_2}}\\
&&[\nabla\times\vec{A}]^\mu=\frac{\partial_\nu A^\lambda}{\sqrt{g_3}}-\frac{\partial_\lambda A^\nu}{\sqrt{g_1}}-\frac{A^\nu}{2(\lambda-\nu)\sqrt{g_1}}-\frac{A^\lambda}{2(\lambda-\nu)\sqrt{g_3}}\\
&&[\nabla\times\vec{A}]^\nu=\frac{\partial_\lambda A^\mu}{\sqrt{g_1}}-\frac{\partial_\mu A^\lambda}{\sqrt{g_2}}+\frac{A^\mu}{2(\lambda-\mu)\sqrt{g_1}}+\frac{A^\lambda}{2(\lambda-\mu)\sqrt{g_2}}
\end{eqnarray}
\subsection{Other coordinate systems}
The Cartesian coordinate system is the trivial one in which the orthonormal basis is identical to the coordinate basis and all commutators are zero. 
 
The spherical coordinate system is one of the more important cases. To illustrate the use of programming codes in obtaining these operators in a very simple procedural way, in  \cite{RAM}  we make available a code to obtain the divergence and curl for the free software Maxima and also for Maple. Moreover, we also include a code to draw general coordinate systems both for Maxima and Maple \cite{RAM}.

 All other coordinate systems are plotted and given in the following Table \ref{tabela}. In this Table some of the coordinate systems are described by Jacobi's elliptic functions. These same coordinate systems could be described by square roots instead, the disadvantage is that it will be necessary a greater amount of charts to cover the Euclidean $E^3$ space. 
\begin{table}
\begin{tabular}{|l|l|l|l|}
\toprule
Parabolic   &  Elliptic   & Parabolic & Paraboloidal\\
cylindrical & cylindrical & &   \\
\toprule
{\footnotesize $ x=\frac{1}{2}\left ( \lambda ^{2}-\mu ^{2}  \right )$ }&{\footnotesize $x=a \cosh \mu  \cos\nu$ }& {\footnotesize$ x= \lambda\mu\cos(\phi)$ }&{\footnotesize $x=d\frac{\mbox{sn}\left (\lambda  ,\kappa  \right )\mbox{sn}\left ( \nu ,\kappa^\prime  \right )}{\mbox{cn}\left ( \lambda ,\kappa  \right )\mbox{cn}\left ( \mu ,\kappa  \right )}$}\\
{\footnotesize $y=\lambda \mu$ }&{\footnotesize  $y=a \sinh \mu \sin\nu$} &{\footnotesize $y= \lambda\mu\sin(\phi)$ }&{\footnotesize $y=d\frac{\mbox{sn}\left (\mu ,\kappa  \right )\mbox{cn}\left ( \nu ,\kappa^\prime  \right )}{\mbox{cn}\left ( \lambda ,\kappa  \right )\mbox{cn}\left ( \mu ,\kappa  \right )}$}\\ 
{\footnotesize $z=z$} &{\footnotesize $z=z$} & {\footnotesize $z= \frac{1}{2}\left ( \lambda^2-\mu^2 \right )$} &{\footnotesize $z=\frac{d}{2}\left [ \frac{\mbox{sn}^{2}\left ( \lambda ,\kappa  \right )}{\mbox{cn}^{2}\left ( \lambda ,\kappa  \right )} - \frac{\mbox{sn}^{2}( \mu ,\kappa )}{\mbox{cn}^2\left ( \mu ,\kappa  \right )}+\frac{\mbox{dn}^{2}( \nu ,\kappa^\prime  )}{k^{'^{2}}}\right ]$}\\
{\scriptsize $\lambda \in R$ and $\mu \in R$} & {\scriptsize $\mu>0$ and $\nu \in [0,2\pi)$ }&{\tiny $\lambda \geq 0$, $\mu \geq 0$, $\phi \in[0,2\pi)$ } &{\scriptsize $b=\kappa a$, $\sqrt{a^2-b^2}=\kappa^\prime a=\sqrt{d}$} \\
& & & {\tiny $\lambda>0$, $\mu \in [-\mu_0,\mu_0]$, $\nu \in [-\nu_0,\nu_0]$} \\
 \raisebox{-\totalheight}{\includegraphics[width=0.2\textwidth, height=20mm]{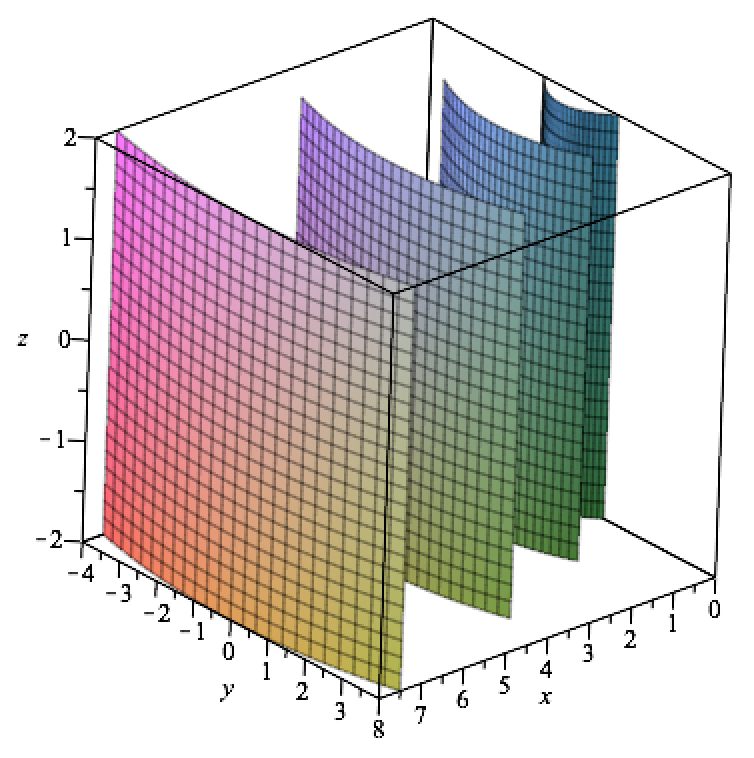}}&  \raisebox{-\totalheight}{\includegraphics[width=0.2\textwidth, height=20mm]{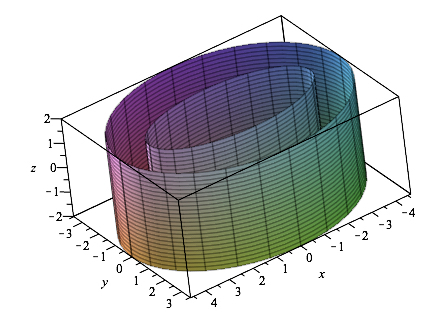}} & \raisebox{-\totalheight}{\includegraphics[width=0.2\textwidth, height=20mm]{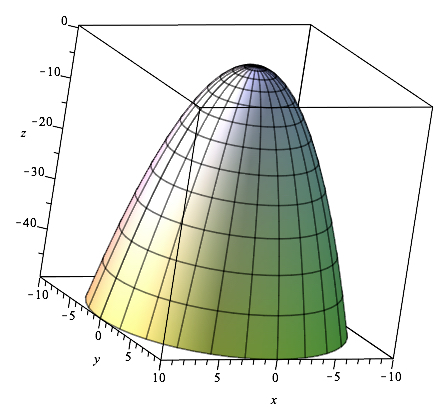}} &\raisebox{-\totalheight}{\includegraphics[width=0.2\textwidth, height=25mm]{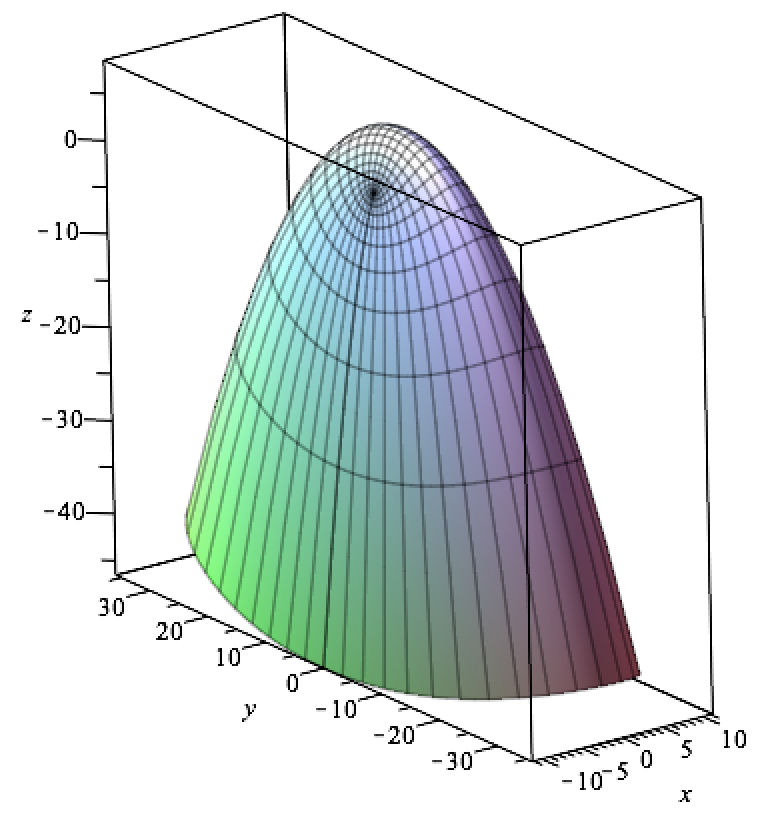}}\\
 \bottomrule
\end{tabular}
\hspace{-1.5cm}
\begin{tabular}{|l|l|l|l|}
\toprule
Conical & Ellipsoidal &  Prolate & Oblate \\ 
\toprule
{\footnotesize $x=r\mbox{dn}\left ( \lambda ,\alpha  \right )\mbox{sn}\left ( \mu ,\beta  \right )$}&{\footnotesize $x=d\frac{\mbox{sn}\left ( \lambda ,\kappa  \right )\mbox{sn}\left ( \mu ,\kappa ^\prime \right )\mbox{dn}\left ( \nu ,\kappa  \right )}{\mbox{cn}\left ( \lambda ,\kappa  \right )}$}&{\scriptsize $ x=d \sinh (\mu)  \sin (\theta) \cos (\phi)$ }&{\scriptsize $x=d \cosh (\mu)  \sin (\theta) \cos (\phi)$}\\
{\footnotesize $y=r\mbox{sn}\left ( \lambda ,\alpha  \right )\mbox{dn}\left ( \mu ,\beta  \right )$}&{\footnotesize $y=d\frac{\mbox{cn}\left ( \mu ,\kappa^\prime  \right )\mbox{cn}\left ( \nu ,\kappa  \right )}{\mbox{cn}\left ( \lambda ,\kappa  \right )}$}&{\scriptsize $y=d \sinh (\mu)  \sin (\theta)\sin (\phi)$}&{\scriptsize $y=d \cosh (\mu)  \sin (\theta)\sin (\phi)$}\\
{\footnotesize $z=r\mbox{cn}\left ( \lambda ,\alpha  \right )\mbox{cn}\left ( \mu ,\beta  \right )$} &{\footnotesize $z=a\frac{\mbox{dn}\left ( \lambda ,\kappa  \right )\mbox{dn}\left ( \mu ,\kappa ^\prime \right )\mbox{sn}\left ( \nu ,\kappa  \right )}{\mbox{cn}\left ( \lambda ,\kappa  \right )}$}& {\scriptsize $z=d \cosh (\mu)  \cos (\theta)$ }&{\scriptsize $z=d \sinh (\mu)  \cos (\theta)$}\\
{\scriptsize $\alpha^2+\beta^2=1$}&{\tiny $b=\kappa a$, $\sqrt{a^2-b^2}=\kappa^\prime a=d$}&{\tiny $\mu>0$, $\theta\in[0,\pi]$,  $\phi\in[0,2\pi)$}&{\tiny $\mu>0$, $\theta \in [0,\pi]$, $\phi \in [0,2\pi]$}\\
{\tiny $r>0$, $\mu \in [-\mu_0,\mu_0]$, } &{\tiny $\lambda>0$, $\mu \in [-\mu_0,\mu_0]$, } & & \\
{\tiny $\lambda \in [-\lambda_0,\lambda_0]$} &{\tiny $\nu \in [-\nu_0,\nu_0]$} & & \\
\raisebox{-\totalheight}{\includegraphics[width=0.2\textwidth, height=20mm]{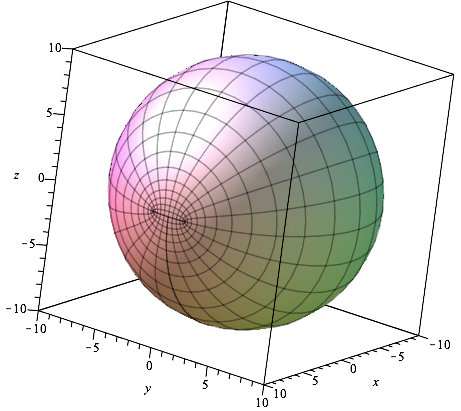}}&\raisebox{-\totalheight}{\includegraphics[width=0.2\textwidth, height=25mm]{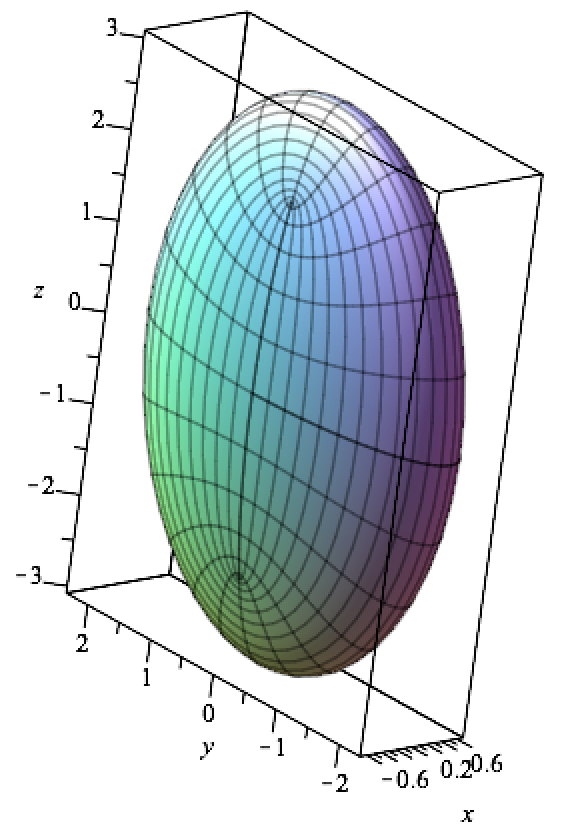}}& \raisebox{-\totalheight}{\includegraphics[width=0.2\textwidth, height=25mm]{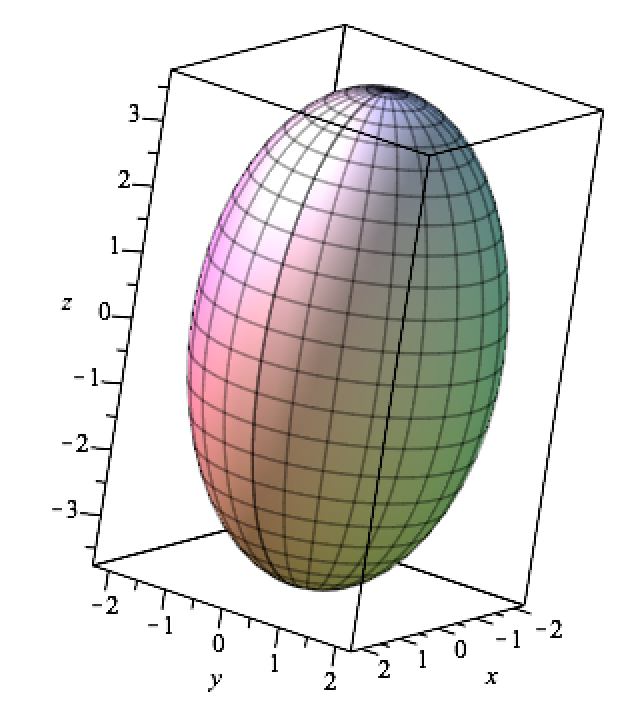}}&\raisebox{-\totalheight}{\includegraphics[width=0.2\textwidth, height=20mm]{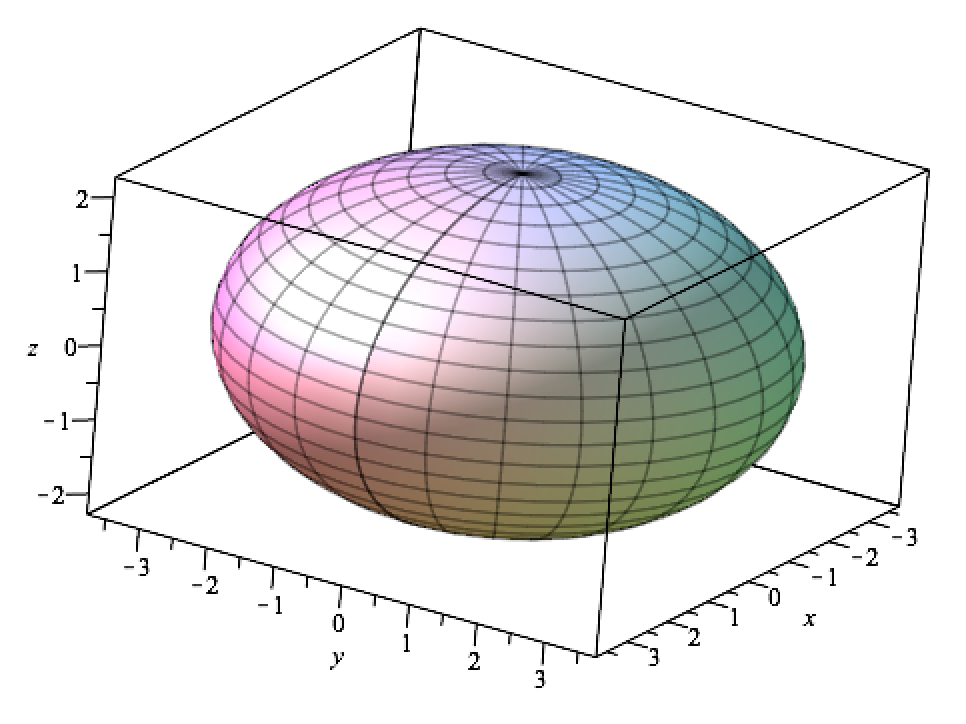}}\\
 \bottomrule
\end{tabular}
\caption{These are the eight coordinate systems other than rectangular, cylindrical and spherical coordinates in which the Laplace operator is separable. It is not difficult to adapt the code given in \cite{RAM} to each one of these coordinate systems or to any other situation of interest, and we leave this exercise to the reader.} \label{tabela}
\end{table}
\newpage
\section{Final remarks}

In the present work we approach the vectorial calculus through the use of the  concept of non coordinate basis, especially the tetrad basis. We present here a practical development to obtain the divergence and curl, that have direct applications in physical problems, for example, in Electromagnetism, Fluid Mechanics, Gravitational Theories and other areas of Physics. The approach is restricted to Euclidean geometry, however, it can be easily extended to curved spaces, which also include situations with gravitational field. 

In this sense, the purpose of this work complements the literature and at the same time introduces sophisticated mathematical concepts with direct applications. It is worth noting that the connection given by (\ref {derivada}) and (\ref {conexao}), could also be obtained through the Fock Ivanenko operator as a particular case when the spin is $1$ \cite{FI}. Therefore, it is presented to the students, either late undergraduate or beginners graduate, a formalism more adapted to differential geometry and theoretical physics.

We thus consider the application of this method for the eleven coordinate systems
mentioned in \cite{Morse_Feshbach}, for which the separation of variables can be applied to Laplace's operator. For each of the eleven coordinates there is a corresponding metric tensor,
and tetrad basis and through the commutator of this basis the connection is obtained. With the appropriate connection, the divergence and curl
for each of these eleven cases can be obtained. We work out  in detail the case for the well known polar coordinates in section \ref{coord.cil}. In \cite{RAM} we show an example code  both for the algebraic manipulator Maple \cite{maple} an the free software Maxima \cite{maxima}, \cite{manuais} to obtain the divergence and curl also for the well known spherical coordinates. Calculations can be made by the reader, either manually, or by adapting the example code to the situation of interest.

\section*{Acknowledgements}
We thank V. Toth in helping us with the Maxima code, and R. A. M\"uller for offering his website to host our codes. We must also thank the Referee for his patience and friendly suggestions.


\end{document}